\documentclass[twocolumn,english,prl,reprint,showpacs]{revtex4-1}
\usepackage[T1]{fontenc}
\usepackage{color}
\usepackage{babel}
\usepackage{verbatim}
\usepackage{amsmath}
\usepackage{amssymb}
\usepackage{graphicx}
\usepackage[unicode=true,pdfusetitle,
 bookmarks=false,
 breaklinks=false,pdfborder={0 0 0},backref=false,colorlinks=true]
 {hyperref}
\begin{document}

\title{Improving the Precision of Weak Measurements by Postselection Measurement}

\author{Shengshi Pang}

\author{Todd A. Brun}

\affiliation{Department of Electrical Engineering, University of Southern California,
Los Angeles, California 90089, USA}
\begin{abstract}
Postselected weak measurement is a useful protocol for amplifying
weak physical effects. However, there has recently been controversy
over whether it gives any advantage in precision. While it is now
clear that retaining failed postselections can yield more Fisher information
than discarding them, the advantage of postselection measurement itself
still remains to be clarified. In this Letter, we address this problem
by studying two widely used estimation strategies: averaging measurement
results, and maximum likelihood estimation, respectively. For the
first strategy, we find a surprising result that squeezed coherent
states of the pointer can give postselected weak measurements a higher
signal-to-noise ratio than standard ones while all standard coherent
states cannot, which suggests that raising the precision of weak measurements
by postselection calls for the presence of ``nonclassicality'' in
the pointer states. For the second strategy, we show that the quantum
Fisher information of postselected weak measurements is generally
larger than that of standard weak measurements, even without using
the failed postselection events, but the gap can be closed with a
proper choice of system state.

\global\long\def\si{|\Phi_{i}\rangle}
\global\long\def\sf{|\Phi_{f}\rangle}
\global\long\def\i{\mathrm{i}}
\global\long\def\e{\mathrm{e}}
\global\long\def\re{\mathrm{Re}}
\global\long\def\im{\mathrm{Im}}
\global\long\def\tr{\mathrm{tr}}
\global\long\def\var{\mathrm{Var}}
\global\long\def\snr{\mathrm{SNR}_{{\rm post}}}
\global\long\def\et{\e^{\i\theta}}
\global\long\def\net{\e^{-\i\theta}}
\global\long\def\snro{\mathrm{SNR}_{{\rm std}}}
\global\long\def\fp{F_{{\rm post}}^{(Q)}}
\global\long\def\fn{F_{{\rm std}}^{(Q)}}
\global\long\def\fa{F_{{\rm all}}^{(Q)}}
\global\long\def\sfi{\langle\Phi_{f}|\Phi_{i}\rangle}
\global\long\def\cf{\langle\Phi_{f}|}
\global\long\def\ci{\langle\Phi_{i}|}
\global\long\def\rdf{\rho_{D}^{\prime}}
\global\long\def\xe{X_{{\rm est}}}
 \global\long\def\avg#1{\langle#1\rangle}

\end{abstract}


\maketitle
\emph{Introduction}.--- Postselected weak measurement is a quantum
measurement protocol first invented by Aharonov, Albert, and Vaidman
in 1988 \cite{Aharonov1988}. It involves weak coupling between the
system and the pointer, but the postselection on the system leads
to a surprisingly counterintuitive effect: the average shift of the
final pointer state can go far beyond the eigenvalue spectrum of the
system observable (multiplied by the coupling constant) in sharp contrast
to the projective quantum measurement. The mechanism behind this effect
is the coherence between the pointer states translated by different
eigenvalues of the system observable, which has an enlightening interpretation
based on superoscillation \cite{Berry2012}.

Postselected weak measurement has aroused enormous research interest
in different fields, due to its ability to amplify tiny physical effects.
Thanks to technical progress in recent years, the weak value has been
measured in experiments \cite{Ritchie1991,Pryde2005,Groen2013,Campagne-Ibarcq2014},
and postselected weak measurements have been applied to measuring
small parameters in various systems, including optical systems \cite{Hosten2008,Dixon2009,Starling2009,Starling2010,Starling2010a,Pfeifer2011,Turner2011,Egan2012,Gorodetski2012,Hofmann2012,Zhou2012,Viza2013,Xu2013,goswami_simultaneous_2014,Magana-Loaiza2014,Mirhosseini2014,viza_experimentally_2015},
atomic systems \cite{Shomroni2013} and NMR \cite{Lu2014}. More experimental
protocols have also been proposed \cite{Brunner2010,Feizpour2011,Li2011,Zilberberg2011,Gotte2012,Nishizawa2012,Wu2012,Dressel2013,Hayat2013,Strubi2013,Zhou2013,Huang2015,Lyons2015}.
A general framework for postselected weak measurement is given in
\cite{Wu2011}, and reviews of the field can be found in \cite{shikano_theory_2012,Kofman2012,Dressel2014}.
Of course, weak value amplification cannot be arbitrarily large in
practice. The condition for the validity of the weak value formalism
was discussed in \cite{Duck1989}, and the limit of amplification
has been studied in \cite{Koike2011,Susa2012,DiLorenzo2014,Pang2014d}.

One of the major goals in postselected weak measurement is to enhance
the sensitivity of estimating small parameters. The experiment of
Starling \emph{et al}. \cite{Starling2009} and the proposal of Feizpour
\emph{et al}. \cite{Feizpour2011} showed that postselection can significantly
raise the signal-to-noise ratio (SNR) of weak measurement. Nevertheless,
some other work has led to a negative conclusion \cite{Knee2013}.
In recent research, it was shown that the failed postselections contain
Fisher information \cite{Tanaka2013,Ferrie2014,Zhang2015}, and even
the distribution probabilities of postselection results can carry
Fisher information \cite{Zhang2015}; thus, discarding postselection
results will generally lead to a loss of precision \cite{Combes2014}. 

To address the issue of low postselection efficiency, Dressel \emph{et
al.} \cite{Dressel2013} and Lyons \emph{et al}. \cite{Lyons2015}
proposed recycling the unpostselected photons to improve the precision.
It was later found \cite{Jordan2014,Pang2014a,Pang2015} that the
successful postselections can concentrate most of the Fisher information
in the pointer, and the Fisher information of postselected weak measurement
can approximately reach the Heisenberg limit \cite{Jordan2014,Pang2014a,Jordan2015,Zhang2015,Pang2015}.
More surprising, weak value amplification can improve the precision
in the presence of technical noise \cite{Feizpour2011,Jordan2014,Pang2015},
and technical noise may increase the SNR of postselected weak measurement
\cite{Kedem2012}. A review of the controversy over the advantage
of weak value amplification is given in \cite{Knee2014a}.

The postselection in a weak measurement includes two steps: first,
measure the system, second, postselect the measurement results. Most
previous research focused on whether failed postselections should
be retained or not, provided the system is measured. However, a more
fundamental problem is whether the system should be measured at all
in order to enhance the precision of weak measurement. If the measurement
on the system could not give any advantage, then it would become meaningless
to study whether the failed postselections should be used or not.
So, this question lies at the heart of postselected weak measurement:
what is the significance of measuring the system in a weak measurement
compared to the standard weak measurement (i.e., without measuring
the system)? Since postselecting and nonpostselecting the results
of measuring the system only lead to a negligible difference in the
Fisher information \cite{Pang2014a,Pang2015}, we will focus only
on comparing postselected weak measurement to the standard weak measurement.

At first glance, this question seems easy to answer: since measuring
the system with proper postselection can amplify the signal, the SNR
can then also be increased. However, the efficiency of postselection
is rather low, which may cancel the benefits of the amplification
effect in the SNR, so the problem becomes subtle. In fact, the numerical
results in \cite{Zhu2011} showed that postselecting the system with
Gaussian pointer states cannot improve the SNR compared with standard
weak measurements, and \cite{Knee2014} found similar results for
the Fisher information of measuring the position or momentum of the
pointer, with the pointer states being real or Gaussian and the weak
values being real or imaginary, respectively.

However, it is important to note that those studies did not optimize
over the choice of the system and pointer states, so they do not rule
out the existence of other choices that may allow postselected weak
measurements to have higher precision. In particular, the Gaussian
states considered heretofore are quite ``classical,'' so it is of
great interest whether using more ``quantum'' states can bring any
advantage for precision. In fact, it has been shown that nonclassical
quantum states can be favorable to some other weak measurement protocols,
e.g., consecutive violations of Clauser-Horne-Shimony-Holt (CHSH)
inequalities \cite{silva_multiple_2015}. Moreover, the measurement
basis of the pointer was not optimized either; hence it is also possible
to have the Fisher information increased by measurements other than
those along the basis of position or momentum of the pointer.

Answering these questions will clarify the advantage of postselection
in weak measurements, and it is exactly the aim of this Letter. We
study the optimal precision of both postselected and standard weak
measurements for general system and pointer states, and investigate
when or whether postselected weak measurements can have higher precision
than standard weak measurements. Moreover, different estimation strategies
may also influence the precision, so we consider two principal estimation
strategies: averaging the measurement results of the pointer (AMR),
and maximum likelihood estimation (MLE), both of which have been widely
used in practice.

For the strategy of AMR, an interesting result we find is that all
standard (i.e., unsqueezed) coherent states do not give weak measurements
an improvement in SNR with postselection, but properly squeezed coherent
states do. This suggests that, for weak value amplification to enhance
the precision, a necessary ingredient is some \emph{nonclassicality}
in the initial pointer states which was missing from previous studies.
This result extends the understanding and feasibility of postselected
weak measurement in parameter estimation.

For the strategy of MLE, we obtain the optimum quantum Fisher information
and show that, even without using the failed postselections, the quantum
Fisher information of postselected weak measurements is generally
higher than that of standard weak measurements.

\emph{Weak value formalism}.--- First, we review the weak value formalism
for postselected weak measurement. Suppose the initial state of the
system is $|\Psi_{i}\rangle$ and the initial state of the pointer
is $|D\rangle$. The interaction Hamiltonian between the system and
the pointer is
\begin{equation}
H_{\mathrm{int}}=gA\otimes\Omega\delta(t-t_{0}),\label{eq:3}
\end{equation}
where the $\delta$ function indicates that the interaction is instantaneous
at time $t_{0}$. Let $\hbar=1$. After the interaction (\ref{eq:3}),
the system is postselected to $\sf$, then the state of the pointer
collapses to $|D_{f}\rangle=\cf\exp(-\i gA\otimes\Omega)\si|D\rangle$
(unnormalized). It can be derived that $|D_{f}\rangle\approx\sfi(1-\i gA_{w}\Omega)|D\rangle$
when $gA_{w}\ll1$, where $A_{w}$ is the \emph{weak value}, defined
as
\begin{equation}
A_{w}=\frac{\cf A\si}{\sfi}.\label{eq:14}
\end{equation}
If one measures an observable $M$ on the pointer state $|D_{f}\rangle$,
it can be obtained \cite{supplemental} that the average shift is
\begin{equation}
\begin{aligned}\langle\Delta M\rangle_{f} & \approx g\im A_{w}(\langle\{\Omega,M\}\rangle_{D}-2\langle\Omega\rangle_{D}\langle M\rangle_{D})\\
 & +\i g\re A_{w}\langle[\Omega,M]\rangle_{D},
\end{aligned}
\label{eq:6}
\end{equation}
where $\langle D|\cdot|D\rangle$ is denoted as $\langle\cdot\rangle_{D}$
for short. And the success probability of postselection is $P_{s}\approx|\sfi|^{2}.$

The weak value (\ref{eq:14}) can be very large when $\sfi\ll1$,
and the dependence of $\langle\Delta M\rangle_{f}$ on $A_{w}$ in
Eq. (\ref{eq:6}) indicates that the average shift can go beyond any
eigenvalue of $A$ in this case. This is the origin of the weak value
amplification.

\emph{Optimal signal-to-noise ratio}.--- First, we study the precision
of postselected weak measurement, then compare it with that of standard
weak measurement, to determine when or whether postselection can assist
weak measurement in precision. 

To quantify the precision of estimating the parameter $g$, a widely
used benchmark is the signal-to-noise ratio of the estimates, defined
as
\begin{equation}
\snr=\frac{\sqrt{NP_{s}}\langle\Delta M\rangle_{f}}{\sqrt{\var(M)_{_{f}}}},\label{eq:12}
\end{equation}
where $N$ is the total number of measurements. The factor $\sqrt{P_{s}}$,
due to $\var(M)_{_{f}}$, scales inversely with the number of successful
postselections. In the first order approximation with respect to $g$,
the spread of the pointer wave function is almost unchanged, so $\var(M)_{_{f}}\approx\var(M)_{D}.$

Note that the quantity defined in Eq. (\ref{eq:12}) is the SNR of
the AMR estimator, not of the measurement results, and it is directly
related to the estimation precision of postselected weak measurement
\cite{supplemental}.

With different pre- and postselections of the system, the SNR is usually
different, so a proper measure for the precision of postselected weak
measurement is the maximum SNR over all possible pre- and postselections.
Direct maximization of the SNR by usual means (such as the variation
method) is rather difficult, since the variation of $\snr$ (\ref{eq:12})
produces a nonlinear equation that is not easy to deal with. 

However, the results of \cite{Pang2014a} offer an alternative possible
approach to this hard problem. In that Letter, the largest success
probability over all postselections of the system for a given weak
value $A_{w}$ was shown to be
\begin{equation}
\max_{\sf}P_{s}\approx\frac{\var(A)_{i}}{\langle A^{2}\rangle_{i}-2\langle A\rangle_{i}\re A_{w}+|A_{w}|^{2}},\label{eq:4}
\end{equation}
where $\langle\cdot\rangle_{i}$ is short for $\ci\cdot\si$. By exploiting
this result, the task of maximizing the SNR over all pre- and postselections
can be simplified to maximizing over all weak values $A_{w}$.

Usually the weak value $A_{w}$ is complex, and can be denoted as
$A_{w}=|A_{w}|\et$, so we can follow a two-step procedure to obtain
the maximum of the $\snr$ over $A_{w}$: first, maximize $\snr$
over $|A_{w}|$, then maximize it over $\theta$. %

The mathematical detail of this optimization is left for the Supplemental
Material. The result of the maximized $\snr$ turns out to be
\begin{equation}
g\eta(\varphi)\sqrt{N\frac{(\langle\{\Omega,M\}\rangle_{D}-2\langle\Omega\rangle_{D}\langle M\rangle_{D})^{2}+|\langle[\Omega,M]\rangle_{D}|^{2}}{\var(M)_{D}}},\label{eq:7}
\end{equation}
where $\varphi=\arctan\frac{\i\langle[\Omega,M]\rangle_{D}}{\langle\{\Omega,M\}\rangle_{D}-2\langle\Omega\rangle_{D}\langle M\rangle_{D}}$
and $\eta(\varphi)=\sqrt{\var(A)_{i}+\langle A\rangle_{i}^{2}\sin^{2}\varphi}$.

In \cite{supplemental}, we also obtained an upper bound on the optimal
$\snr$ based on (\ref{eq:7}).

\emph{When can SNR be increased?}--- The maximum SNR (\ref{eq:7})
quantifies the metrological performance of postselected weak measurement.
To address the question of when (or whether) postselection can improve
the SNR of weak measurement, we need to further compare (\ref{eq:7})
with the maximum SNR of standard weak measurement.

Before proceeding with this question, it is helpful to note that in
the average shift of the pointer (\ref{eq:6}), the real part of the
weak value is assigned with the commutator between $\Omega,\,M$ and
the imaginary part with the covariance between $\Omega,\,M$. These
coefficients can be quite large with proper pointer states and will
not be counterbalanced by the low postselection probability while
the weak values may be. So it opens the possibility of increasing
the SNR by postselection.

In a standard weak measurement, the average shift in the observable
$M$ on the postinteraction pointer state is $\langle\Delta M\rangle=\i g\langle A\rangle_{i}\langle[\Omega,M]\rangle_{D}$
\cite{supplemental}, and $\max\langle A\rangle_{i}=\lambda_{\max}(A)$,
so the optimal SNR is
\begin{equation}
\max\snro=g\frac{\sqrt{N}|\lambda_{\max}(A)\langle[\Omega,M]\rangle_{D}|}{\sqrt{\var(M)_{D}}}.\label{eq:13}
\end{equation}
The ratio between the optimal SNR of postselected and standard weak
measurements is, therefore,
\begin{equation}
s=\frac{\sqrt{\var(A)_{i}\csc^{2}\varphi+\langle A\rangle_{i}^{2}}}{|\lambda_{\max}(A)|}.\label{eq:1}
\end{equation}

Obviously, since $\csc^{2}\varphi\geq1$, when $\si\rightarrow|\lambda_{\max}(A)\rangle$,
$\sqrt{\var(A)_{i}\csc^{2}\varphi+\langle A\rangle_{i}^{2}}\geq|\lambda_{\max}(A)|$,
and thus $s\geq1,$ which means that postselection in weak measurement
will not reduce the SNR at least, but this is still not enough. The
key question is when (or whether) $\csc^{2}\varphi>1$ can hold, so
that postselection gives an increase of the SNR compared with standard
weak measurement.

To answer this question, we move to Fock space. Suppose that $\Omega=q,\,M=p$.
Then $[\Omega,M]{}_{D}=\i.$ In Fock space, $q$ and $p$ can be represented
by $q=(a+a^{\dagger})/\sqrt{2},\,p=(a-a^{\dagger})/\sqrt{2}\i,$ so
$\{q,p\}=\i(a^{\dagger2}-a^{2}),$ and $\csc^{2}\varphi=1+\left|\langle a^{\dagger2}\rangle+\langle a\rangle_{D}^{2}-\langle a^{\dagger}\rangle_{D}^{2}-\langle a^{2}\rangle_{D}\right|^{2}.$

When the initial pointer state $|D\rangle$ is a standard coherent
state, $\csc^{2}\varphi=1$, so standard coherent states cannot give
postselected weak measurements any advantage in SNR over standard
weak measurements. This generalizes the results of \cite{Zhu2011,Knee2014},
and suggests that ``classical'' pointer states are not able to improve
the SNR of postselected weak measurements.

An interesting question is whether introducing ``nonclassicality''
to the pointer state can ``activate'' the advantage of postselected
weak measurement in SNR. Consider squeezed coherent states for the
pointer. Suppose the initial state $|D\rangle$ of the pointer is
\begin{equation}
|\xi,\alpha\rangle=\exp\frac{1}{2}(\xi^{*}a^{2}-\xi a^{\dagger2})|\alpha\rangle,\label{eq:10}
\end{equation}
where $\xi$ is the squeeze parameter. Let $\xi=r\e^{\i\theta}$,
then one can find \cite{supplemental}
\begin{equation}
\csc^{2}\varphi=1+4(\sin\theta\sinh r\cosh r)^{2}.\label{eq:11}
\end{equation}

It is clear from (\ref{eq:11}) that when $\sin\theta\neq0$, one
can acquire $\csc^{2}\varphi>1$ with a large $r$, so according to
(\ref{eq:1}), if $\var(A)_{i}\neq0$, the SNR of postselected weak
measurements exceeds that of standard weak measurements in this case.
This shows that nonclassicality really can assist the postselection
to improve the SNR of weak measurements. It is in a similar spirit
to Ref. \cite{Feizpour2011}: correlations, classical or quantum,
can increase the SNR of weak measurements.

\begin{figure}
\includegraphics[scale=0.6]{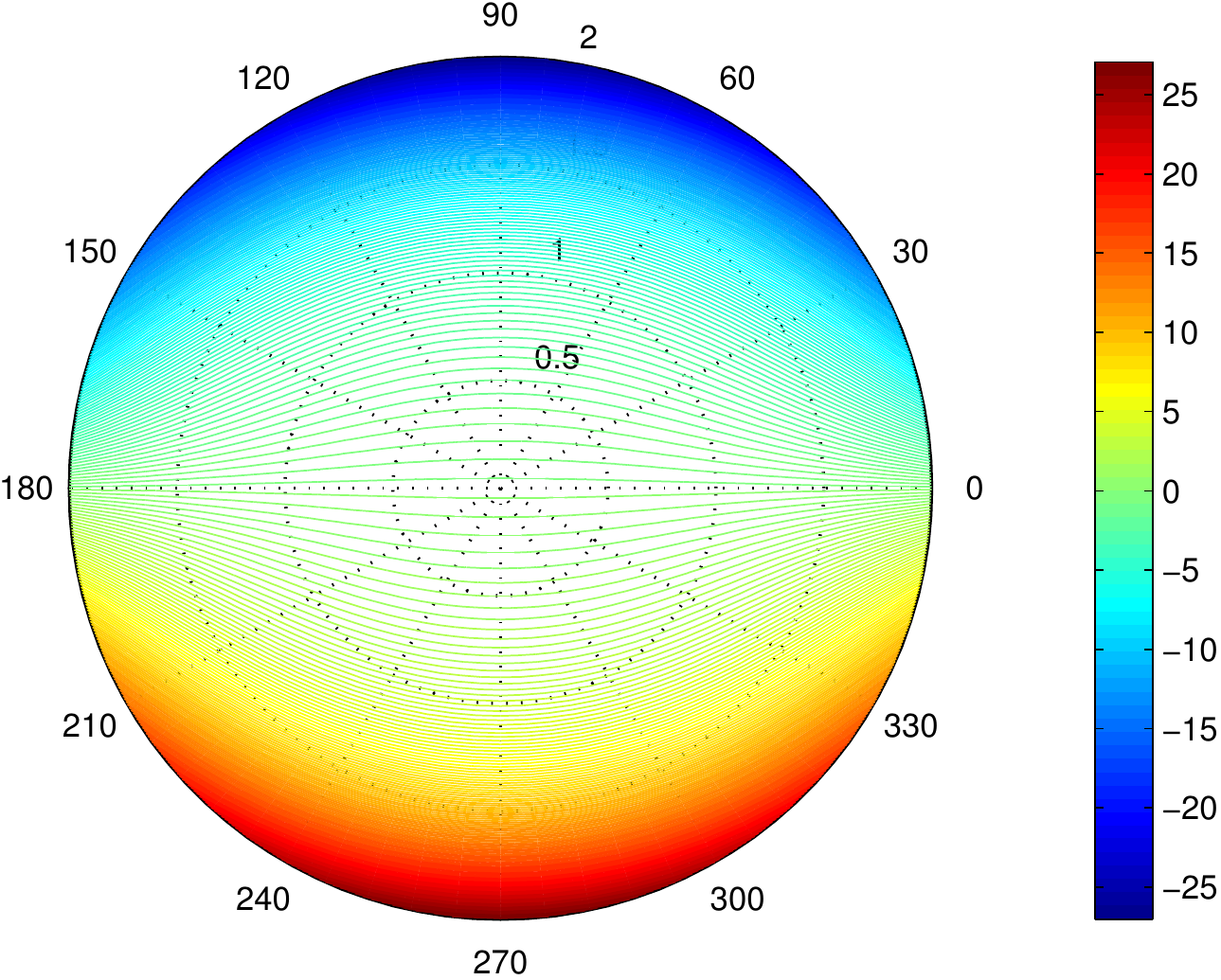}

\caption{(Color online) The contours of $s$ are plotted for the squeezed vacuum
states $|\xi,0\rangle=\exp\frac{1}{2}(\xi^{*}a^{2}-\xi a^{\dagger2})|0\rangle$
with $|\xi|\leq2$ and $|\arg\xi|\leq\pi$. The interaction Hamiltonian
is $g\sigma_{z}\otimes q$ with $g=10^{-5}$. The weak value is fixed
to $20\protect\i$. The momentum $p$ is measured on the pointer after
the postselection. Each point in the figure represents a $\xi$ on
the complex plane, and the color indicates the corresponding value
of $s$, which is the ratio between the SNR of postselected and standard
weak measurements. It clearly shows $|s|$ can be much larger than
$1$ with proper $\xi$, implying an increase in the SNR by postselecting
the system. The sign of $s$ denotes the relative sign between the
results of postselected and standard weak measurements.\label{fig:snr}}
\end{figure}

To illustrate the above result, Fig. \ref{fig:snr} plots the contours
of the ratio $s$ on the complex plane of $\xi$ for the squeezed
vacuum state $|\xi,0\rangle$. Improvement of SNR can be explicitly
observed in the figure.

Why are squeezed coherent states more beneficial to the SNR than standard
coherent states? It can be roughly understood from the following.
The SNR of postselected weak measurement can be shown to be bounded
by $\sqrt{\var(\Omega)_{D}}$ \cite{supplemental}, and the SNR of
standard weak measurement is proportional to $1/\sqrt{\var(M)_{D}}$
(see Eq. (\ref{eq:13})). The ratio between them is approximately
$\sqrt{\var(\Omega)_{D}\var(M)_{D}}$. Since coherent states have
minimal uncertainty, $\sqrt{\var(\Omega)_{D}\var(M)_{D}}$ does not
change and keeps the minimum for conjugate quadratures $\Omega$ and
$M$. In contrast, squeezing can increase $\var(\Omega)_{D}$ and
decrease $\var(M)_{D}$, so it can simultaneously increase the SNR
of both types of weak measurements. However, squeezed coherent states
no longer have the minimum uncertainty, so $\sqrt{\var(\Omega)_{D}\var(M)_{D}}$
can be increased. Hence, the SNR of postselected weak measurement
can be raised more than that of standard weak measurement.

It is worth noting that squeezing may also simultaneously decrease
$\var(\Omega)_{D}$ and increase $\var(M)_{D}$ instead, and the SNR
of postselected weak measurement can still be higher than that of
standard weak measurement. But in this case, the SNR of both types
of weak measurements are decreased, so it should be avoided in practice.

\emph{Optimal quantum Fisher information}.--- Next, we turn to the
precision of weak measurements using maximum likelihood estimation
strategy. Once again, our goal is to determine whether postselected
or standard weak measurement has greater precision, and what conditions
determine the advantage.

The exact variance of the MLE estimator is usually difficult to obtain;
however, Cram\'{e}r and Rao \cite{Cramer1946} showed that it is inversely
bounded by the Fisher information, and this bound can be saturated
in the asymptotic limit. So we will use Fisher information as the
measure of precision for MLE instead.

As different measurements on the pointer produce different Fisher
information, a proper benchmark for the precision of MLE is the maximum
Fisher information over all possible measurements on the pointer,
called the \emph{quantum Fisher information} \cite{Braunstein1994,Braunstein1996},
and it gives a more general bound than that found by working in only
one specific measurement basis. For a pure $g$-dependent state $|\psi_{g}\rangle$,
the quantum Fisher information of estimating $g$ is $F^{(Q)}=4(\langle\partial_{g}\psi_{g}|\partial_{g}\psi_{g}\rangle-|\langle\psi_{g}|\partial_{g}\psi_{g}\rangle|^{2})$.

In a postselected weak measurement, the pointer state after postselecting
the system is $|D_{f}\rangle\approx\e^{-\i gA_{w}\Omega}|D\rangle$,
so $|\partial_{g}D_{f}\rangle\approx-(\i A_{w}\Omega+\langle\Omega\rangle_{D}\im A_{w})|D\rangle$,
and the quantum Fisher information is approximately \cite{supplemental}
\begin{equation}
\fp\approx4P_{s}|A_{w}|^{2}\var(\Omega)_{D},
\end{equation}
where we note the dependence on the postselection probability $P_{s}$.
The maximum $P_{s}$ is given by (\ref{eq:4}), therefore, the maximum
quantum Fisher information over all postselections given the weak
value $A_{w}$ is
\begin{equation}
\fp\approx\frac{4|A_{w}|^{2}\var(A)_{i}\var(\Omega)_{D}}{\langle A^{2}\rangle_{i}-2\langle A\rangle_{i}\re A_{w}+|A_{w}|^{2}}.\label{eq:5}
\end{equation}

Now, the task is just to maximize $\fp$ over $A_{w}$. This maximization
can be achieved by a two-step procedure similar to maximizing $\snr$
\cite{supplemental}, and the result is
\begin{equation}
\max\fp\approx4\langle A^{2}\rangle_{i}\var(\Omega)_{D}.\label{eq:2}
\end{equation}

As a comparison, consider the standard weak measurement. In this case,
the postinteraction pointer state is generally a mixed state since
the pointer is entangled with the system by the weak interaction.
The quantum Fisher information for mixed states is much more complex
than that for pure states, and a general analytical result is unavailable. 

However, with the weak coupling limit $gA_{w}\ll1$, this difficulty
can be significantly reduced, since the postinteraction pointer state
can be approximated to a pure state $|D_{f}\rangle\approx\e^{-\i g\langle A\rangle_{i}\Omega}|D\rangle$
\cite{supplemental}. Then, one can immediately derive the quantum
Fisher information for standard weak measurement
\begin{equation}
\fn\approx4\langle A\rangle_{i}^{2}\var(\Omega)_{D}.\label{eq:0}
\end{equation}
Now, comparing $\fn$ with $\fp$, the ratio between them can be obtained:
\begin{equation}
\frac{\fp}{\fn}\approx\frac{\langle A^{2}\rangle_{i}}{\langle A\rangle_{i}^{2}}.\label{eq:15}
\end{equation}

The result (\ref{eq:15}) compares the \emph{quantum} Fisher information
between postselected and standard weak measurements for \emph{every}
possible state of the system, in contrast to Ref. \cite{Knee2014,Knee2014a}
where the Fisher information of measuring the pointer along the position
or momentum basis was compared between the two types of weak measurements
for their respective optimal system states (with additional assumptions
as reviewed in the Introduction). Eq. (\ref{eq:15}) indicates that
the initial state of the system decides the ratio of quantum Fisher
information, and implies the postselected weak measurement generally
possesses more Fisher information than the standard weak measurement,
except that the latter can catch up when the initial system is in
an eigenstate of $A$.

Ref. \cite{Tanaka2013,Ferrie2014,Zhang2015} made the comparison between
using and discarding failed postselections, given that the system
is measured. Ref. \cite{Jordan2014,Pang2014a,Pang2015} showed that
the difference between the Fisher information in these two cases can
be shrunk to be negligibly small. Combining (\ref{eq:15}) with those
results, if we denote the quantum Fisher information retaining all
failed postselections as $\fa$, then
\begin{equation}
\fa\gtrsim\fp\geq\fn.\label{eq:9}
\end{equation}
This clearly shows the relation of the quantum Fisher information
between different types of weak measurements, and clarifies when the
postselected weak measurement has metrological advantage. The first
inequality of (\ref{eq:9}) reflects the results of \cite{Tanaka2013,Ferrie2014,Jordan2014,Pang2014a,Zhang2015,Pang2015},
and the equality sign of the second inequality accords with \cite{Knee2014,Knee2014a}.

\emph{Remark}.--- The results for SNR and Fisher information at first
glance seem quite different: a significant advantage can be given
by postselected weak measurements over standard weak measurements
in SNR, while the advantage is quite limited in Fisher information.
The difference is rooted in the performances of the two estimators
behind them, AMR and MLE, respectively. MLE has the minimum variance
over all estimators, while AMR does not, and the Fisher information
is usually an upper bound on the precision of MLE (except for Gaussian
distributions) which can be achieved only asymptotically. Because
of these differences, the SNR has more room to be improved than the
Fisher information by optimizing the measurement strategy and the
initial states of the system and pointer. These results indicate that
the advantage of postselected weak measurements has dependence on
the choice of estimation strategy.

This research was supported by the ARO MURI under Grant No. W911NF-11-1-0268.

\bibliographystyle{apsrev4-1}
\bibliography{optimalsnr}

\newpage \setcounter{equation}{0} \renewcommand{\theequation}{S\arabic{equation}} \onecolumngrid \setcounter{enumiv}{0}

\section*{Supplemental Material}

\section{Average shift of the pointer}

In this section, we show how to derive the average shift of the pointer
for postselected weak measurements and standard weak measurements,
respectively.

\subsection{Postselected weak measurements}

The interaction Hamiltonian between the system and the pointer is
\begin{equation}
H_{\mathrm{int}}=gA\otimes\Omega\delta(t-t_{0}).\label{eq:12-1}
\end{equation}
Suppose the initial state of the system is $\si$, and the initial
state of the pointer is $|D\rangle$. The system is postselected in
the state $\sf$ after the interaction (\ref{eq:12-1}), and the pointer
collapses to the state (unnormalized)
\begin{equation}
|D_{f}\rangle=\cf\exp(-\i gA\otimes\Omega)\si|D\rangle.
\end{equation}

When $g$ is sufficiently small, $|D_{f}\rangle$ is approximately
\begin{equation}
\begin{aligned}|D_{f}\rangle & \approx\cf(1-\i gA\otimes\Omega)\si|D\rangle\\
 & =\langle\Psi_{f}|\Psi_{i}\rangle(1-\i gA_{w}\Omega)|D\rangle,
\end{aligned}
\label{eq:11-1}
\end{equation}
where $A_{w}$ is the so-called \emph{weak value}, defined as $A_{w}=\frac{\langle\Psi_{f}|A|\Psi_{i}\rangle}{\langle\Psi_{f}|\Psi_{i}\rangle}.$
The success probability of postselection is
\begin{equation}
P_{s}\approx|\langle\Psi_{f}|\Psi_{i}\rangle|^{2}.
\end{equation}

The average shift in the measured value of the pointer observable
$M$ is
\begin{equation}
\langle\Delta M\rangle_{f}=\frac{\langle D_{f}|M|D_{f}\rangle}{\langle D_{f}|D_{f}\rangle}-\langle M\rangle_{D}.\label{eq:6-1}
\end{equation}
We denote $\langle D|\cdot|D\rangle$ as $\langle\cdot\rangle_{D}$
for short throughout the paper. Note that
\begin{equation}
\begin{aligned}\langle D_{f}|M|D_{f}\rangle & \approx|\langle\Psi_{f}|\Psi_{i}\rangle|^{2}(\langle M\rangle_{D}+\i g\re A_{w}\langle[\Omega,M]\rangle_{D}+g\im A_{w}\langle\{\Omega,M\}\rangle_{D}),\\
\langle D_{f}|D_{f}\rangle & \approx|\langle\Psi_{f}|\Psi_{i}\rangle|^{2}(1+2g\im A_{w}\langle\Omega\rangle_{D}).
\end{aligned}
\label{eq:3-1}
\end{equation}
By plugging (\ref{eq:3-1}) into (\ref{eq:6-1}), we see that
\begin{equation}
\begin{aligned}\langle\Delta M\rangle_{f} & =g\im A_{w}(\langle\{\Omega,M\}\rangle_{D}-2\langle\Omega\rangle_{D}\langle M\rangle_{D})\\
 & +\i g\re A_{w}\langle[\Omega,M]\rangle_{D},
\end{aligned}
\end{equation}
which is Eq. (3) in the main text.

\subsection{Standard weak measurements}

In a standard weak measurement, the interaction between the system
and pointer is also given by (\ref{eq:12-1}), but there is no measurement
on the system, so the post-interaction system-pointer state is $\exp(-\i gA\otimes\Omega)\si|D\rangle,$
and the average shift of the pointer is
\begin{equation}
\langle\Delta M\rangle_{f}=\langle\Phi_{i}|\langle D|\exp(\i gA\otimes\Omega)M\exp(-\i gA\otimes\Omega)\si|D\rangle-\langle M\rangle_{D}.
\end{equation}

When $g\ll1$,
\begin{equation}
\begin{aligned}\exp(\i gA\otimes\Omega)M\exp(-\i gA\otimes\Omega) & \approx(I+\i gA\otimes\Omega)M(I-\i gA\otimes\Omega)\\
 & \approx M+\i gA\otimes[\Omega,M].
\end{aligned}
\end{equation}
Therefore,
\begin{equation}
\langle\Delta M\rangle_{f}=\i g\langle A\rangle_{i}\langle[\Omega,M]\rangle_{D}.\label{eq:14-1}
\end{equation}

\section{Optimum signal-to-noise ratio of postselected weak measurements}

In this section, we detail the the optimization of the signal-to-noise
ratio (SNR) over the weak value $A_{w}$ for postselected weak measurements
by the two-step procedure outlined in the main text. We will also
obtain an upper bound on the maximum SNR as a by-product.

The SNR of the estimated value of $g$ by a postselected weak measurement
with interaction (\ref{eq:12-1}) is
\begin{equation}
\snr=\frac{\sqrt{NP_{s}}\langle\Delta M\rangle_{f}}{\sqrt{\var(M)_{_{f}}}}.\label{eq:17}
\end{equation}
where $M$ is the observable we measure on the pointer.

Since the numerator of $\snr$ is $O(g)$, we need only approximate
the denominator of $\snr$ to $O(1)$ to guarantee that the $\snr$
has a precision of $O(g)$. Thus, we can assume $\var(M)_{_{f}}\approx\var(M)_{D}.$

\subsection{Relation between SNR and parameter uncertainty}

Before proceeding to obtain the optimum SNR, we first clarify the
relation between the SNR defined in (\ref{eq:17}) and the uncertainty
in the parameter to estimate. 

Note that weak value amplification is a \emph{linear} amplification,
$\langle\Delta M\rangle_{f}=gc$, where $c=\im A_{w}(\langle\{\Omega,M\}\rangle_{D}-2\langle\Omega\rangle_{D}\langle M\rangle_{D})+\i\re A_{w}\langle[\Omega,M]\rangle_{D}$
is the amplification factor (see Eq.~(3) of the main text), and $c$
is $\partial_{g}\langle\Delta M\rangle_{f}$. So $\snr$ can be rewritten
as 
\begin{equation}
\snr=\frac{g}{\frac{\sqrt{\var(M)_{_{f}}/N_{s}}}{\partial_{g}\langle\Delta M\rangle_{f}}},\label{eq:1-1}
\end{equation}
where $N_{s}=NP_{s}$ and $\var(M)_{_{f}}/N_{s}$ is the variance
of $\langle M\rangle_{f}$. Since $\langle\Delta M\rangle_{f}=\langle M\rangle_{f}-\langle M\rangle_{i}$
where $\langle M\rangle_{i}$ is a constant, $\sqrt{\var(M)_{_{f}}/NP_{s}}$
is also the variance of $\langle\Delta M\rangle_{f}$.

Generally, the deviation of an estimator $\xe$ from the real value
of a parameter $X$ \cite{Braunstein1994} is defined as 
\begin{equation}
\delta X=\frac{\xe}{|\partial_{X}\langle\xe\rangle_{X}|}-X.
\end{equation}
and the standard deviation of the estimator $\xe$ is 
\begin{equation}
\sqrt{\avg{\delta X^{2}}}=\frac{\sqrt{\var(\xe)}}{|\partial_{X}\langle\xe\rangle_{X}|},\label{eq:2-1}
\end{equation}
where we assume the estimator $\xe$ to be unbiased and linear in
the parameter. (If the estimator is biased, an additional bias term
will appear on the right side of (\ref{eq:2-1}), which corresponds
to the bias of the estimator.)

The estimator $\xe$ for the SNR defined in (\ref{eq:17}) is the
average of the measurement results, i.e., $\xe=(m_{1}+\cdots+m_{N_{s}})/N_{s}$,
where $m_{1},\cdots,m_{N_{s}}$ are the outputs of measuring $M$
on the pointer state, and the expectation value of the estimator $\xe$
is $\langle\Delta M\rangle_{f}$. Note that $\var(\xe)=\var(X)/N_{s}$,
then by comparing Eqs. (\ref{eq:1-1}) and (\ref{eq:2-1}), one can
immediately see that $\snr$ is equal to the parameter $g$ divided
by the variance of the esitmator. So $\snr$ is exactly proportional
to the inverse of the variance of the estimator, which is the uncertainty
of the parameter in the estimation. Therefore, the $\snr$ in (\ref{eq:17})
is a well defined measure of the parameter uncertainty.

\subsection{Maximum SNR}

The postselection probability $P_{s}$ can take different values for
a given weak value $A_{w}$ by varying the pre- and postselections.
However, the maximum $P_{s}$ given the weak value $A_{w}$ \cite{Pang2014a}
is
\begin{equation}
\max P_{s}\approx\frac{\var(A)_{i}}{\langle A^{2}\rangle_{i}-2\langle A\rangle_{i}\re A_{w}+|A_{w}|^{2}}.\label{eq:4-1}
\end{equation}
Thus, the $\snr$ can be written as
\begin{equation}
\snr=g\sqrt{N\frac{\var(A)_{i}}{\var(M)_{D}}}\frac{\im A_{w}(\langle\{\Omega,M\}\rangle_{D}-2\langle\Omega\rangle_{D}\langle M\rangle_{D})+\i\re A_{w}\langle[\Omega,M]\rangle_{D}}{\sqrt{\langle A^{2}\rangle_{i}-2\langle A\rangle_{i}\re A_{w}+|A_{w}|^{2}}}.\label{eq:13-1}
\end{equation}

Now, let $A_{w}=|A_{w}|\et$. We can maximize $\snr$ over $A_{w}$
by a two-step procedure: first, maximize $\snr$ over $|A_{w}|$,
then over $\theta$. We first maximize $\snr$ over $|A_{w}|$. Note
that (\ref{eq:13-1}) can be written
\begin{equation}
\snr=g\sqrt{N\frac{\var(A)_{i}}{\var(M)_{D}}}\frac{(\langle\{\Omega,M\}\rangle_{D}-2\langle\Omega\rangle_{D}\langle M\rangle_{D})\sin\theta+\i\langle[\Omega,M]\rangle_{D}\cos\theta}{\sqrt{\langle A^{2}\rangle_{i}|A_{w}|^{-2}-2\langle A\rangle_{i}\cos\theta|A_{w}|^{-1}+1}}.
\end{equation}
The minimum of $\langle A^{2}\rangle_{i}|A_{w}|^{-2}-2\langle A\rangle_{i}\cos\theta|A_{w}|^{-1}+1$
is
\begin{equation}
1-\frac{\langle A\rangle_{i}^{2}\cos^{2}\theta}{\langle A^{2}\rangle_{i}},
\end{equation}
if $\cos\theta>0$, and the critical point of $|A_{w}|$ is
\begin{equation}
|A_{w}|_{c}=\frac{\langle A^{2}\rangle_{i}}{\langle A\rangle_{i}\cos\theta}.\label{eq:8-1}
\end{equation}
So, the maximum of $\snr$ over $|A_{w}|$ is
\begin{equation}
\max_{|A_{w}|}\snr=g\sqrt{N\frac{\var(A)_{i}\langle A^{2}\rangle_{i}}{\var(M)_{D}}}\frac{(\langle\{\Omega,M\}\rangle_{D}-2\langle\Omega\rangle_{D}\langle M\rangle_{D})\sin\theta+\i\langle[\Omega,M]\rangle_{D}\cos\theta}{\sqrt{\langle A^{2}\rangle_{i}-\langle A\rangle_{i}^{2}\cos^{2}\theta}}.\label{eq:2-1}
\end{equation}

Next, we maximize (\ref{eq:2-1}) over $\theta$. For simplicity,
let
\begin{equation}
\varphi=\arctan\frac{\i\langle[\Omega,M]\rangle_{D}}{\langle\{\Omega,M\}\rangle_{D}-2\langle\Omega\rangle_{D}\langle M\rangle_{D}}.\label{eq:15-1}
\end{equation}
Then (\ref{eq:2-1}) can be simplified to
\begin{equation}
\max_{|A_{w}|}\snr=gK(\theta)\sqrt{N\frac{\var(A)_{i}\langle A^{2}\rangle_{i}}{\var(M)_{D}}}\sqrt{(\langle\{\Omega,M\}\rangle_{D}-2\langle\Omega\rangle_{D}\langle M\rangle_{D})^{2}+|\langle[\Omega,M]\rangle_{D}|^{2}},
\end{equation}
where
\begin{equation}
K(\theta)=\frac{\sin(\theta+\varphi)}{\sqrt{\langle A^{2}\rangle_{i}-\langle A\rangle_{i}^{2}\cos^{2}\theta}}.
\end{equation}
The key is to maximize $K(\theta)$ over $\theta$ to obtain the maximum
SNR. Note that $K(\theta)$ can be rewritten as
\begin{equation}
\begin{aligned}K(\theta) & =\frac{\sin(\theta+\varphi)}{\sqrt{\var(A)_{i}+\langle A\rangle_{i}^{2}(\sin(\theta+\varphi)\cos\varphi-\cos(\theta+\varphi)\sin\varphi)^{2}}}\\
 & =\frac{1}{\sqrt{(\var(A)_{i}+\langle A\rangle_{i}^{2}\cos^{2}\varphi)-2\langle A\rangle_{i}^{2}\sin\varphi\cos\varphi\cot(\theta+\varphi)+(\var(A)_{i}+\langle A\rangle_{i}^{2}\sin^{2}\varphi)\cot^{2}(\theta+\varphi)}}.
\end{aligned}
\end{equation}

It is obvious that $K(\theta)$ is maximized when
\begin{equation}
\cot(\theta+\varphi)=\frac{\langle A\rangle_{i}^{2}\sin\varphi\cos\varphi}{\var(A)_{i}+\langle A\rangle_{i}^{2}\sin^{2}\varphi},
\end{equation}
and the maximum of $K(\theta)$ is
\begin{equation}
\begin{aligned}\max_{\theta}K(\theta) & =\frac{1}{{\displaystyle \sqrt{\var(A)_{i}+\langle A\rangle_{i}^{2}\cos^{2}\varphi-\frac{\langle A\rangle_{i}^{4}\sin^{2}\varphi\cos^{2}\varphi}{\var(A)_{i}+\langle A\rangle_{i}^{2}\sin^{2}\varphi}}}}\\
 & =\sqrt{\frac{\var(A)_{i}+\langle A\rangle_{i}^{2}\sin^{2}\varphi}{\var(A)_{i}\langle A^{2}\rangle_{i}}}.
\end{aligned}
\end{equation}
Therefore, the maximum of the SNR over $A_{w}$ finally turns out
to be
\begin{equation}
\max\snr=g\eta(\varphi)\sqrt{N\frac{(\langle\{\Omega,M\}\rangle_{D}-2\langle\Omega\rangle_{D}\langle M\rangle_{D})^{2}+|\langle[\Omega,M]\rangle_{D}|^{2}}{\var(M)_{D}}},\label{eq:7-1}
\end{equation}
where
\begin{equation}
\eta(\varphi)=\sqrt{\var(A)_{i}+\langle A\rangle_{i}^{2}\sin^{2}\varphi}.\label{eq:16}
\end{equation}

\subsection{Upper bound of SNR}

The Robertson-Schr\"odinger uncertainty relation tells that for any
two Hermitian operators $\Omega_{1}$ and $\Omega_{2}$, there is
\begin{equation}
\Delta\Omega_{1}\Delta\Omega_{2}\geq\frac{1}{2}\sqrt{|\langle[\Omega_{1},\Omega_{2}]\rangle|^{2}+|\langle\{\Omega_{1},\Omega_{2}\}\rangle-2\langle\Omega_{1}\rangle\langle\Omega_{2}\rangle|^{2}}.
\end{equation}
By this inequality, an upper bound on $\snr$ can be immediately obtained
from (\ref{eq:7-1}):
\begin{equation}
\max\snr\leq2g\eta(\varphi)\sqrt{N\var(\Omega)_{D}}.\label{eq:1-1}
\end{equation}

\section{Signal-to-noise ratio with different pointer states}

\subsection{Standard coherent state for the pointer}

According to (\ref{eq:14-1}), the SNR of a standard weak measurement
is
\begin{equation}
\snro=g\frac{\sqrt{N}|\langle A\rangle_{i}\langle[\Omega,M]\rangle_{D}|}{\sqrt{\var(M)_{D}}}.
\end{equation}
Since $\max\langle A\rangle_{i}=\lambda_{\max}(A)$, the optimal SNR
is
\begin{equation}
\max\snro=g\frac{\sqrt{N}|\lambda_{\max}(A)\langle[\Omega,M]\rangle_{D}|}{\sqrt{\var(M)_{D}}}.
\end{equation}
The ratio between the SNR of postselected and standard weak measurements
is
\begin{equation}
s=\frac{\max\snr}{\max\snro}=\frac{\eta(\varphi)}{\lambda_{\max}(A)}\sqrt{1+\frac{(\langle\{\Omega,M\}\rangle_{D}-2\langle\Omega\rangle_{D}\langle M\rangle_{D})^{2}}{|\langle[\Omega,M]\rangle_{D}|^{2}}}.
\end{equation}

According to the definition of $\varphi$ in (\ref{eq:15-1}),
\begin{equation}
\frac{(\langle\{\Omega,M\}\rangle_{D}-2\langle\Omega\rangle_{D}\langle M\rangle_{D})^{2}}{|\langle[\Omega,M]\rangle_{D}|^{2}}=\cot^{2}\varphi,
\end{equation}
so by using (\ref{eq:16}), $s$ can be simplified to
\begin{equation}
s=\frac{\sqrt{\var(A)_{i}\csc^{2}\varphi+\langle A\rangle_{i}^{2}}}{|\lambda_{\max}(A)|}.
\end{equation}

Since $\var(A)_{i}=\langle A^{2}\rangle_{i}-\langle A\rangle_{i}^{2}$
and $\csc^{2}\varphi\geq1$, we have $\var(A)_{i}\csc^{2}\varphi+\langle A\rangle_{i}^{2}\geq\langle A^{2}\rangle_{i}$,
so when $\si\rightarrow|\lambda_{\max}(A)\rangle$, $s\geq1.$ The
key question is when is $\csc^{2}\varphi>1$, so that $s>1$? 

Suppose $\Omega=q,\,M=p$. Then $\langle[\Omega,M]\rangle_{D}=\i$,
and
\begin{equation}
\csc^{2}\varphi=1+(\langle\{q,p\}\rangle_{D}-2\langle q\rangle_{D}\langle p\rangle_{D})^{2}.
\end{equation}
In the Fock space, $q$ and $p$ can be represented by $q=\frac{1}{\sqrt{2}}(a+a^{\dagger}),\,p=\frac{1}{\sqrt{2}\i}(a-a^{\dagger}),$
so $\{q,p\}=\i(a^{\dagger2}-a^{2}),$ and
\begin{equation}
\csc^{2}\varphi=1-(\langle a^{\dagger2}-a^{2}\rangle_{D}+\langle a\rangle_{D}^{2}-\langle a^{\dagger}\rangle_{D}^{2})^{2}.
\end{equation}

When the initial pointer state $|D\rangle$ is a standard coherent
state, $\langle a^{\dagger2}\rangle_{D}=\langle a^{\dagger}\rangle^{2},\,\langle a^{2}\rangle_{D}=\langle a\rangle^{2},$
so $\csc^{2}\varphi=1$, which means that standard coherent states
of the pointer cannot give postselected weak measurements any advantage
in SNR compared with standard weak measurements.

\subsection{Squeezed coherent state for the pointer}

Now turn to squeezed coherent states for the pointer. Suppose,
\begin{equation}
|D\rangle=|\xi,\alpha\rangle=\exp\frac{1}{2}(\xi^{*}a^{2}-\xi a^{\dagger2})|\alpha\rangle,\label{eq:10-1}
\end{equation}
where $\xi=r\e^{\i\theta}$ is the squeeze parameter. Then it can
be shown that
\begin{equation}
\begin{aligned}\langle a\rangle_{D} & =\alpha\cosh r-\alpha^{*}\e^{\i\theta}\sinh r,\\
\langle a^{\dagger}\rangle_{D} & =\langle a\rangle_{D}^{*},\\
\langle a^{2}\rangle_{D} & =\alpha^{2}\cosh^{2}r+\alpha^{*2}\e^{2\i\theta}\sinh^{2}r-2|\alpha|^{2}\e^{\i\theta}\sinh r\cosh r-\e^{\i\theta}\sinh r\cosh r,\\
\langle a^{\dagger2}\rangle_{D} & =\langle a^{2}\rangle_{D}^{*}.
\end{aligned}
\end{equation}
So in this case,
\begin{equation}
\csc^{2}\varphi=1+4(\sin\theta\sinh r\cosh r)^{2}.\label{eq:9-1}
\end{equation}

When $r>0$ and $\sin\theta\neq0$, then $\csc^{2}\varphi>1$. Therefore
the SNR of postselected weak measurements in this case can be enhanced
beyond the SNR of standard weak measurements. This implies that properly
squeezed coherent states can increase the SNR of weak measurements
by postselection while standard coherent states cannot.

\section{Optimum quantum Fisher information}

In this section, we obtain the quantum Fisher information for postselected
weak measurements and standard weak measurements, respectively.

\subsection{The case of postselected weak measurements}

For a pure parameter-dependent state $|\psi_{g}\rangle$, the quantum
Fisher information of estimating $g$ is \cite{Braunstein1994,Braunstein1996}
\begin{equation}
F^{(Q)}=4(\langle\partial_{g}\psi_{g}|\partial_{g}\psi_{g}\rangle-|\langle\psi_{g}|\partial_{g}\psi_{g}\rangle|^{2}).\label{eq:5-1}
\end{equation}

We first consider postselected weak measurements. According to (\ref{eq:11-1})
and (\ref{eq:3-1}), the pointer state after the postselection on
the system in a postselected weak measurement is
\begin{equation}
|D_{f}\rangle\approx\frac{\e^{-\i gA_{w}\Omega}|D\rangle}{\sqrt{1+2g\im A_{w}\langle D|\Omega|D\rangle}},
\end{equation}
so
\begin{equation}
|\partial_{g}D_{f}\rangle\approx-(\i A_{w}\Omega+\langle\Omega\rangle_{D}\im A_{w})|D\rangle.
\end{equation}
Therefore,
\begin{equation}
\begin{aligned}\langle\partial_{g}D_{f}|\partial_{g}D_{f}\rangle & \approx|A_{w}|^{2}\langle\Omega^{2}\rangle_{D}-\im^{2}A_{w}\langle\Omega\rangle_{D}^{2},\\
\langle D_{f}|\partial_{g}D_{f}\rangle & \approx-\i\re A_{w}\langle\Omega\rangle_{D}.
\end{aligned}
\end{equation}
Hence, the quantum Fisher information of the post-interaction pointer
state in this type of weak measurements is
\begin{equation}
\fp\approx4P_{s}|A_{w}|^{2}\var(\Omega)_{D}.
\end{equation}

Now, we just need to maximize $\fp$ over $A_{w}$. This maximization
can be achieved by a two-step procedure similar to that for maximizing
$\snr$. The optimal $|A_{w}|$ is still given by (\ref{eq:8-1}),
so the maximum $\fp$ over $|A_{w}|$ is
\begin{equation}
\max_{|A_{w}|}\fp\approx\frac{4\langle A^{2}\rangle_{i}\var(A)_{i}\var(\Omega)_{D}}{\langle A^{2}\rangle_{i}-\langle A\rangle_{i}^{2}\cos^{2}\theta}.
\end{equation}
Obviously, when $\cos\theta=\pm1,$ $\fp$ is maximized, so the global
maximum of $\fp$ is
\begin{equation}
\max\fp\approx4\langle A^{2}\rangle_{i}\var(\Omega)_{D}.
\end{equation}

\subsection{The case of standard weak measurements}

Next, we consider standard weak measurements. In a standard weak measurement,
the post-interaction pointer state is usually a mixed state, since
the pointer is entangled with the system by the interaction. The reduced
density matrix of the pointer after the interaction is
\begin{equation}
\rdf=\tr_{S}(\exp(-\i gA\otimes\Omega)\si|D\rangle)\langle\Phi_{i}|\langle D|\exp(\i gA\otimes\Omega)).
\end{equation}
Suppose the eigenstates of $A$ are $|a_{i}\rangle,\i=1,\cdots,d$,
where the $a_{i}$'s are the corresponding eigenvalues, and the initial
state of the system is $\si=\sum_{k}\alpha_{k}|a_{k}\rangle$. Then
$\rdf$ becomes
\begin{equation}
\rdf=\sum_{k}|\alpha_{k}|^{2}\exp(-\i ga_{k}\Omega)|D\rangle\langle D|\exp(\i ga_{k}\Omega).
\end{equation}

Since $g\ll1$, $\exp(-\i ga_{k}\Omega)\approx1-\i ga_{k}\Omega$,
and
\begin{equation}
\begin{aligned}\rdf & \approx\sum_{k}|\alpha_{k}|^{2}(1-\i ga_{k}\Omega)|D\rangle\langle D|(1+\i ga_{k}\Omega)\\
 & \approx\sum_{k}|\alpha_{k}|^{2}(|D\rangle\langle D|-\i ga_{k}[\Omega,|D\rangle\langle D|])\\
 & \approx|D\rangle\langle D|-\i g\langle A\rangle_{i}[\Omega,|D\rangle\langle D|],
\end{aligned}
\end{equation}
where we have used $\sum_{k}|\alpha_{k}|^{2}=1$ and $\sum_{k}|\alpha_{k}|^{2}a_{k}=\langle A\rangle_{i}$.

Again, because $g\ll1$,
\begin{equation}
|D\rangle\langle D|-\i g\langle A\rangle_{i}[\Omega,|D\rangle\langle D|\approx\exp(-\i g\langle A\rangle_{i}\Omega)|D\rangle\langle D|\exp(\i g\langle A\rangle_{i}\Omega).
\end{equation}
After the interaction, the pointer is approximately in a pure state:
\begin{equation}
|D_{f}\rangle\approx\exp(-\i g\langle A\rangle_{i}\Omega)|D\rangle.
\end{equation}
From Eq. (\ref{eq:5-1}), we can immediately derive the quantum Fisher
information of the post-interaction pointer state:
\begin{equation}
\fn\approx4\langle A\rangle_{i}^{2}\var(\Omega)_{D}.\label{eq:0-1}
\end{equation}

\end{document}